\documentclass[12pt]{article}
\usepackage{latexsym}
\def\bbbr{{\rm I\!R}}     
\def\bbbp{{\rm I\!P}}     
\def\bbbc{{\rm I\!\!\! C}}  
\def\bbbz{{\bf  Z}}       
\def\bbbn{{\bf  N}}       
\def\bbbO{{\cal O}}     
\def\bbbL{{\cal L}}     
\newtheorem{definition}{Definition}
\newtheorem{th1}{Theorem}[section]
\newtheorem{pr}[th1]{Proposition}

\newtheorem{cor}[th1]{Corollary}

\def\hfleche#1#2{\smash{\mathop{\vbox{\hbox to
12mm{{\rightarrowfill}}}}\limits^{#1}_{#2}}} 

\def\vfleche#1#2{\llap{$\scriptstyle  #1$}\left\downarrow \vbox
to 6mm{}\right.\rlap{$\scriptstyle #2$}}

\def\diag{\def\normalbaselines{\baselineskip25pt\lineskip3pt
                \lineskiplimit3pt}%
                \matrix}

\begin{document}

\title{ Generalized Jacobians of spectral curves and completely integrable systems}
\author{Lubomir Gavrilov\\
\normalsize \it Laboratoire Emile Picard, UMR 5580,
 Universit\'e Paul Sabatier\\
\normalsize \it 118, route de Narbonne,
 31062 Toulouse Cedex, France }
\date{September 1996}
\maketitle


\section{Introduction}

Let $M^J$ be the affine vector space of all complex matrix polynomials $A(x)$ in a variable
$x$, of fixed degree $d$ and dimension $r$
$$
A(x)= J x^d+A_{d-1}x^{d-1}+...+A_0 , \; A_i \in {\bf gl}_r(\bbbc) 
$$
where $J\in {\bf gl}_r(\bbbc)$ is  a fixed 
matrix. The matricial polynomial Lax equations
\begin{equation}
\label{axa}
\frac{d}{dt} A(x)= [\frac{A^k(a)}{x-a},A(x)], k\in \bbbn, a\in \bbbc
\end{equation}
are well known to be Hamiltonian (with respect to several compatible Poisson 
structures on $M^J$)
and completely integrable. The corresponding Hamiltonian vector fields define
a complete set of commuting vector fields on the isospectral manifolds
$$ M_P^J= \{A(x)\in M^J: det(A(x)-y I_r)= P(x,y) \} .$$
The system  (\ref{axa}) has an obvious symmetry group 
 $G= \bbbp{\bf GL}_r(\bbbc;J)$ which is the subgroup of the projective group
$\bbbp{\bf GL}_r(\bbbc)$ formed by matrices which commute with $J$. 
The group $G$  acts on $M^J$ by conjugation,
 the
action is Poisson, and the reduced Hamiltonian system is completely integrable
 too. As the symmetry group $G$ acts freely and properly on the general 
isospectral manifold $M_P^J$, then $M_P^J$ can be considered as the
total space
 of
a holomorphic principal fibre bundle $\xi$
with base $M_P^J /G$, structural group $G$, and natural projection map
$$
M_P^J \stackrel{\phi}{\rightarrow} M_P^J /G .
$$

The purpose of the present article is to describe the algebraic structure of 
the above fibre bundle. 
Our main result, Theorem \ref{gav}, implies that when the spectral curve
$X$ defined by $\{(x,y)\in \bbbc^2: P(x,y)=0\}$ is smooth,
then $M_P^J$ is smooth and bi-holomorphic to a Zariski open subset of
the generalized Jacobian variety
$J(X')$. 
The curve $X'$ is singular and as a topological space it is just $X$
with its ``infinite" points $\infty_1,\infty_2,...,\infty_r$ identified to a 
single point $\infty$. Thus $J(X')$ is a non-compact commutative algebraic
group and it can be
described as an extension of the usual Jacobian $J(X)$ by the algebraic group
$G=(\bbbc^*)^{s-1}\times \bbbc^{r-s}$, where $s\leq r$ is the number of distinct
eigenvalues of the leading term $J$
\begin{equation}
\label{aaa}
0\stackrel{ }{\rightarrow} G
 \stackrel{}{\rightarrow}
J(X')\stackrel{\phi}{\rightarrow}
J(X) \rightarrow 0 \; .
\end{equation}
As analytic spaces $J(X')$ and $J(X)$ are complex tori
$$
J(X')=\bbbc^{p_a}/\Lambda', \;J(X)=\bbbc^{p_g}/\Lambda
$$
where $\Lambda', \Lambda$ are lattices of rank $2p_g+s-1$ and $2p_g$ respectively,
$p_g$ is the genus of $X$, and $p_a=p_g+r-1$ is the arithmetic genus of $X'$.
The generalized Jacobian $J(X')$ can be also considered as 
the total space of
a holomorphic principal fibre bundle 
with base $J(X)$, projection  $\phi$, and structural group $G$. 
The group $G$ is then identified with the symmetry group $\bbbp{\bf GL}_r(\bbbc;J)$
of (\ref{axa}), and the  manifold $M_P^J/G$ with a Zariski
open subset of the usual Jacobian $J(X)=J(X')/G$. The algebraic 
description of the reduced invariant manifold $M_P^J/G$ is a well
known  result proved by
 A.Beauville \cite{Beauville} and 
M.R.Adams, J.Harnad, J.Hurtubise \cite{Adams} (see also 
 M. Adler, P. van Moerbeke \cite{Adler}, and 
section 8.2. of the survey \cite{Reyman} by A.G.Reyman and 
M.A. Semenov-Tian-Shansky ).

The Hamiltonian structures of the differential equation (\ref{axa}) is briefly
recalled in section \ref{integrable} where we show that the Hamiltonian vector fields
(\ref{axa}) for $k\in\bbbn, a\in \bbbc$ define translation invariant vector fields
on the 
generalized Jacobian $J(X')$, so the system is algebraically completely integrable.

In the case when the spectral curve $X$ is singular, the variety
$M_P^J/\bbbp{\bf GL}_r(\bbbc;J)$ was  studied by 
M.R.Adams, J.Harnad, J.Hurtubise \cite{Adams} (see also 
Beauville \cite[p.218]{Beauville}, 
P. van Moerbeke and D. Mumford \cite[p.112]{Moerbeke}). Note that our
approach is quite the opposite in the sense that, while in \cite{Adams} the
singular spectral curve $X$ is desingularized, in the present article 
the regular spectral curve $X$ is  singularized  to a curve $X'$. 

We conclude the paper with two applications of Theorem \ref{gav} (section \ref{Examples}).
We prove that even in the simplest case when $X$ is elliptic and $G=\bbbc^*$,
the extension (\ref{aaa}) is not trivial, and then describe 
the corresponding
two degrees of freedom algebraically completely integrable system. It turns out 
to be the well known symmetric (Lagrange) top, and $\bbbc^*$ is just 
the complexified  group of rotations about the symmetry axis of the
top. This result, proved {\em ad hoc} by 
Gavrilov and Zhivkov \cite{Gav}, motivated the present paper.
 Another classical problem
 related to Theorem \ref{gav} is to solve a system of hyperelliptic
differential equations (Jacobi \cite{Jacobi}, 1846). We prove that
the phase space of such a system is the generalized Jacobian
$J(X')$ of a hyperelliptic
curve $X$ with two points at ``infinity" identified, 
each orbit is a straight line isomorphic to $\bbbc^*$, and the space of orbits
 is parameterized by the usual Jacobian $J(X)$. This gives a new proof of Jacobi's 
theorem.

\section{ Spectral curves and their Jacobians }
\label{general}

A polynomial
$$
P(x,y)= y^r+s_1(x)y^{r-1} + ... + s_r(x)
$$
 is called spectral, provided that 
the affine curve $\{(x,y)\in \bbbc^2: P(x,y)=0\}$ is the spectrum
of some polynomial $r \times r$ matrix $A(x)$
$$
P(x,y)= det(A(x) - y.I_r) \; .
$$
In this case $deg(s_i(x)) \leq i.d$, where $d$ is the degree of $A(x)$
\begin{equation}
\label{ax}
A(x)= A_d x^d+A_{d-1}x^{d-1}+...+A_0 , \; A_i \in {\bf gl}_r(\bbbc) \; .
\end{equation}
Consider the weighted projective space 
$\bbbp^2(d)=\bbbc^3 \backslash \{0\} / \bbbc^* $, where the $\bbbc^*$-action on
$\bbbc^3$ is defined by
$$
t\cdot (x,y,z) \rightarrow (tx,t^dy,tz), \; t\in \bbbc^* \; .
$$
$\bbbp^2(d)$ is a compact complex surface with one singular point $\{[0,1,0]\}=  \bbbp^2(d)_{sing}$.
 The affine curve  $\{(x,y)\in \bbbc^2: det(A(x) - y.I_r) = 0 \}$
is naturally embedded in $\bbbp^2(d)$, 
$$
\bbbc^2 \rightarrow \bbbp^2(d) : (x,y) \mapsto [x,y,1] ,
$$
and the condition $deg(s_i(x)) \leq i.d$ shows that its closure $X$ is contained 
in the smooth surface
$\bbbp^2(d)_{reg}=\bbbp^2(d)\backslash \{[0,1,0]\}$.
Let $x$ be an affine coordinate on $\bbbp^1$.
The surface $\bbbp^2(d)_{reg}$ is identified with the total space of the
holomorphic line bundle $\bbbO_{\bbbp^1}(d)$ with base $\bbbp^1$ and 
 projection 
$$
\pi :\bbbp^2(d)_{reg} \rightarrow \bbbp^1: [x,y,z] \rightarrow [x,z] .
$$
 The induced projection
\begin{equation}
\label{pi}
\pi : X \rightarrow \bbbp^1
\end{equation}
is a ramified covering of degree $r$, and over the affine plane $\bbbc$ it is simply the first projection
$$
\pi: \{(x,y)\in \bbbc^2 : P(x,y) = 0 \} \rightarrow \bbbc : (x,y) \rightarrow x \; .
$$
\begin{definition} (spectral curve of $A(x)$)
We define the spectral curve $X$ of the matrix polynomial $A(x)$ (\ref{ax}),  to be the closure
of the affine curve $\{(x,y)\in \bbbc^2: det(A(x) - y.I_r) = 0 \}$ 
in the total space of the line bundle $\bbbO_{\bbbp^1}(d)$.
\end{definition}

From now on we fix the spectral polynomial $P(x,y)$ and  suppose that the spectral curve 
$X$ is {\it smooth} and irreducible.

We are going now to singularize the  curve $X$.
Let  $m = \sum_{i=1}^s n_i P_i$, 
$P_i \in X, n_i >0$, be an effective divisor on $X$. 
To the pair $(X, m)$  we associate a singular curve $X'= X_{reg}\cup \infty$, where
if  $S= \cup_{i=1}^s P_i$ is the support of $m$, then $X_{reg}= X-S$,  and $\infty$ is a single point. 
 The structure sheaf 
$\bbbO'$ of $X'\sim (X,m)$ is defined in the following way. Let $\bbbO_{X'}$ be the direct image of 
the structure sheaf $\bbbO=\bbbO_X$ under the canonical projection $X \rightarrow X'$.
Then
$$
\bbbO'_P = 
\left\{
\begin{array}{l}
\bbbO_P, \; \; P\in X_{reg} \\
\bbbc + i_\infty , \; \; P= \infty 
\end{array} 
\right.
$$
where  $i_\infty$ is the ideal of $\bbbO_\infty$ formed by the 
functions $f$ having a zero at $P_i$ of order at least $n_i$. 
Thus a regular function $f$ on $X'$ is a 
regular function $f$ on $X$, and such that for some $c\in \bbbc$ and any $i$ holds
$v_{P_i}(f-c) \geq n_i$, where $v_P(.)$ is the order function. If $p_g$ is the genus of $X$ then
the arithmetic genus $p_a$ of the singular curve $X'$ is $p_a=p_g+deg(m)-1$.

{\it Example} Let $m= P^++P^-$ be a divisor on the Riemann surface $X$. 
Then in a 
neighborhood of $\infty$ the singularized curve $X'$ is analytically isomorphic
either to the germ of analytical curve $xy=0$ ($P^+\neq P^-$), or to 
$y^2=x^3$ ($P^+=P^-$).

\begin{definition} (singularized spectral curve of $A(x)$)
\label{m}
If $\pi$ is the projection (\ref{pi}) and
$\infty =[1,0] \in \bbbp^1$ the ``infinite" divisor, then the effective divisor
$m=\pi^* (\infty) $
is called 
modulus of the spectral curve $X$.
We have $deg(m)=r$ and we denote by $X'$  the singular curve associated to the regular curve $X$
and to the modulus $m$.
\end{definition}

{\em Remark} In the Serre's book \cite{Serre} any affective divisor $m$ on
a regular algebraic curve $X$ is called modulus. Indeed, the moduli space of
singularized curves $X'\sim (X,m)$, $p_g(X)=const.$, $deg(m)=const.$, is of
dimension strictly bigger  than the dimension of the 
moduli space of regular curves $X$.

We shall  recall now the construction of the generalized Jacobian variety 
$J(X')$ of a singular 
curve $X' \sim (X,m)$. 
For proofs we refer the reader to Serre \cite{Serre}.

A holomorphic line bundle $L'$ on $X'$ is described by an open covering $\{ U_\alpha \}_\alpha$ of
$X'$ and transition functions 
$g_{\alpha\beta} \in \bbbO'^*(U_\alpha  \cap U_\beta )$, such that
$$
g_{\alpha\beta}\cdot g_{\beta\alpha}=1, \;
 g_{\alpha\beta} \cdot g_{\beta\gamma} \cdot g_{\gamma\alpha}=1 \;.
$$
Two line bundles $L'_1,L'_2$ on $X'$ are equivalent if and only if there exist 
$f_\alpha \in \bbbO'^*(U_\alpha)$ such that $g_{\alpha\beta}^1= (f_\alpha/f_\beta) \cdot g_{\alpha\beta}^2$.
Thus the Picard group $Pic(X')$ of equivalence classes of holomorphic line bundles on the singular curve 
$X'$ is just  $H^1(X',\bbbO'^*)$. If $D$ is a divisor on  $X_{reg}$ with local equations 
$\{f_\alpha\}$, then the functions 
$g_{\alpha\beta}=f_\alpha/f_\beta \in \bbbO'^*(U_\alpha  \cap U_\beta ) $ define a line bundle
$L'_D$ on $X'$, and any holomorphic line bundle on $X'$ can be written in such a way. 
Two line bundles $L'_{D_1}, L'_{D_2}$ on $X'$ a equivalent if and only if $D_1 \stackrel{m}{\sim} D_2$. This means
that there exists a global meromorphic function $f$ on $X$, such that $(f)=D_1-D_2$ and
$v_{P_i}(f-1) \geq n_i$, $i=1,2...s$.
Let $\bbbL(D)$ be the sheaf of sections of the holomorphic line bundle $L_D$ over the smooth curve
$X$ and suppose as before that the support of $D$ is contained in the set of regular points $X_{reg}$.
 Then
the sheaf of sections $\bbbL'(D)$ of the line bundle $L'_D$ over $X'$ is defined as
$$
\bbbL'(D)_P = 
\left\{
\begin{array}{c}
\bbbO'_\infty, \; \; P= \infty \\
\bbbL(D)_P, \; \; P \neq \infty \; . 
\end{array} 
\right.
$$

The sheaf $\bbbL'(D)$is a locally free $\bbbO'$ module  of rank one, that is to say it is invertible.
More generally, there is an one-to-one correspondence between isomorphism classes of 
invertible sheaves on $X'$, and isomorphism classes of line bundles over $X'$.
This set of isomorphism classes is a group under the operation $\otimes$, 
$\bbbL'(D_1)\otimes \bbbL'(D_2) = \bbbL'(D_1+D_2)$, called Picard group $Pic(X')$ of the curve $X'$.
Let $Pic^0(X')$ be the subgroup of $Pic(X')$ formed by
degree zero line bundles. It is called {\it Jacobian} of $X'$ and we denote $J(X')=Pic^0(X')$. The Jacobian
$J(X')$ of the singular algebraic curve $X'$ has a natural structure of commutative algebraic group.
As an analytic manifold we have
$$
J(X')=  H^0(X,\Omega ^1( m))^*/H_1(X_{reg},\bbbz) = \bbbc^{p_a}/ \Lambda' \; ,
$$
where  $\Lambda'$ is a rank $2p_g+s-1$ lattice, and $\Omega ^1( m)$ is the sheaf
of meromorphic one-forms $\omega $, such that $(\omega ) \geq - m$. 
Similarly, for the usual Jacobian $J(X)=Pic^0(X) \subset J(X') $, we have
$$
J(X)= H^0(X,\Omega ^1)^*/H_1(X,\bbbz) = \bbbc^{p_g}/ \Lambda \; ,
$$
where $\Lambda \subset \Lambda'$ is a rank $2p_g$ lattice.
$J(X')$ is a non-trivial {\it extension} of
$J(X)$ by the algebraic group $G = (\bbbc^*)^{s-1} \times \bbbc^{deg(m)-s}$
\begin{equation}
\label{extension}
0\stackrel{ }{\rightarrow} G
 \stackrel{ }{\rightarrow}
J(X')\stackrel{\phi}{\rightarrow}
J(X) \rightarrow 0
\end{equation}
where $\phi(\bbbL'(D))=\bbbL(D)$. 
This means that the sequence is exact in the usual sense
and moreover the algebraic structure of $G$ (respectively of $J(X)$) is induced
(respectively quotient) of the algebraic structure of $J(X')$.
Both $J(X)$ and $J(X')$ are commutative algebraic groups. Note
however that $J(X')$ is non-compact. Indeed, while the topological space of $J(X)$
is $(S^1)^{2p_g}$, the one of $J(X')$ is $(S^1)^{2p_g+s-1}\times \bbbr^{2deg(m)-s-1}$.
To every extension (\ref{extension}) we associate a holomorphic
principal fibre bundle with total space $J(X')$, base $J(X)$, projection
$\phi$, and structural group $G=(\bbbc^*)^{s-1} \times \bbbc^{deg(m)-s}$. 
Two extensions are equivalent if and only if the associated principal bundles
are equivalent.

Let $J^{p_a}(X)=Pic^{p_a}(X)$ be the variety (isomorphic to the Jacobian $J(X)$)
formed by line bundles of degree $p_a=p_g+r-1$ on $X$, and denote by
$J^{p_a}(X')=Pic^{p_a}(X')$ the variety (isomorphic to the generalized 
Jacobian $J(X')$)
formed by line bundles of degree $p_a=p_g+r-1$ on the singularized curve $X'$.
Denote further by $\Theta$ the canonical Riemann theta divisor of $J^{p_a}(X)$
formed by special line bundles $L_D$
of degree $p_a=p_g+r-1$, that is to say $dim H^1(X,\bbbL(D)) \neq 0$. By Riemann-Roch
theorem
$$
dim H^0(X,\bbbL(D))=deg(D)-p_g+1 + dim H^1(X,\bbbL(D))= r+ dim H^1(X,\bbbL(D))
$$
so $\Theta$ is the set of line bundles $L_D$ with at least $r+1$ holomorphic
sections. Similarly, let $\Theta' \subset J^{p_a}(X')$
be the canonical divisor
formed by degree $p_a$ special line bundles $L'(D)$,  
that is to say $dim H^1(X',\bbbL'(D)) \neq 0$. By Riemann-Roch theorem
$$
dim H^0(X',\bbbL'(D))=deg(D)-p_a+1 + dim H^1(X',\bbbL'(D))= 1+ dim H^1(X',\bbbL'(D))
$$
so such bundles have at least two holomorphic sections. It is easy to see that
$\Theta'= \phi^{-1} (\Theta)$, where $\phi$ is the map induced by (\ref{extension}).

Let $M_P$ be the variety of $r\times r$ polynomial matrices of degree $d$ (\ref{ax}), which have a
fixed spectral polynomial $P(x,y)$
$$ M_P= \{A(x): det(A(x)-y I_r)= P(x,y) \} .$$
and let  $M_P^J=M_P \cap M^J$ be the isospectral manifold formed by matrices
 of the form (\ref{ax})
 with fixed leading term
$A_d=J$
\begin{equation}
\label{axj}
A(x)= J x^d+A_{d-1}x^{d-1}+...+A_0 , \; A_i \in {\bf gl}_r(\bbbc) \; .
\end{equation}
The stabilizer 
$$
\bbbp{\bf GL}_r(\bbbc;J) = \{ R \in \bbbp{\bf GL}_r(\bbbc): RJR^{-1}=J \} 
$$
 of 
$\bbbp{\bf GL}_r(\bbbc)$ at $J\in {\bf gl}_r(\bbbc)$ 
is a commutative algebraic group isomorphic to 
$(\bbbc^*)^{s-1} \times \bbbc^{deg(m)-s}$. It is a well known fact that
$M_P^J$ is a smooth manifold, $\bbbp{\bf GL}_r(\bbbc;J)$ 
acts freely and properly 
on $M_P^J$  by conjugation, and the quotient space $M_P^J/\bbbp{\bf GL}_r(\bbbc;J)$ 
is a smooth manifold  biholomorphic to $J(X)-\Theta$
\cite{Adams,Beauville}.
Consider the holomorphic principal fibre bundle $\xi$ with total space
$M_P^J$, structural group $\bbbp{\bf GL}_r(\bbbc;J)$,
base $M_P^J/\bbbp{\bf GL}_r(\bbbc;J)$, and natural projection map
$\varphi : M_P^J \rightarrow M_P^J/\bbbp{\bf GL}_r(\bbbc;J)$.
Consider also the associate principal bundle $\eta$ with base space $J(X)-\Theta$, total space
$J(X')-\Theta'$, projection map $\phi$, and structural group $G$ 
(see (\ref{extension})).

The main result of the present paper is the following
\begin{th1}
\label{gav}
The holomorphic principal bundles $\xi$ and $\eta$ are isomorphic. In particular the isospectral
manifold $M_P^J$ is smooth and bi-holomorphic to the Zariski open subset
 $J(X')-\Theta'$ of the generalized Jacobian $J(X')$ of the singularized 
spectral curve $X'$.
\end{th1}

We may resume Theorem \ref{gav} in the following commutative diagram
\begin{equation}
\label{commute}
\diag{
&&&&M_P^J & \hfleche{\varphi}{} & M_P^J/\bbbp{\bf GL}_r(\bbbc;J) && \cr
&&&&\vfleche{}{l'}&  &\vfleche{}{l} && \cr
&&&&J(X')-\Theta' & \hfleche{\phi}{} &  J(X)-\Theta  && \cr
&&&&\vfleche{}{}&  &\vfleche{}{} && \cr
0 & \hfleche{}{} & G & \hfleche{}{} & J(X') & \hfleche{}{} &   J(X)  & \hfleche{}{} & 0  \cr}
\end{equation}
in which the maps $l, l'$ are biholomorphic,
$$l': \varphi^{-1}(b) \rightarrow \phi^{-1} \circ l(b)$$
is an isomorphism of algebraic groups for every 
$b\in  M_P^J/\bbbp{\bf GL}_r(\bbbc;J)  $, and
the exact sequence in (\ref{commute})
 is an  extension of the algebraic
group $J(X)$ by $G$.

\paragraph{Proof of Theorem \ref{gav}} 
Recall that a matrix $B\in {\bf GL}_r(\bbbc)$ is called {\it regular}
if one of the following equivalent conditions is satisfied

\begin{description}
\item[-]  all  eigenspaces of $B$ are of dimension one
\item[-] the minimal and the  characteristic polynomials of $B$ are equal
\item[-] the variety ${\bf GL}_r(\bbbc;B)$ is of dimension $r$
\end{description}

We shall use the following  

\begin{pr}
\label{regular}
If the spectral curve $X$ of the matrix polynomial $A(x)$ is smooth, then $A(x)$ is
regular for any fixed $x\in \bbbp^1$ (for $x=\infty$ this means that
the leading term $A_d$ of $A(x)$ is regular).
\end{pr}

Indeed, if for some $x_0\in \bbbc$ the matrix 
$A(x_0) = (a_{ij}(x_0))_{i,j}$  is not regular, 
then there exists $y_0$ such that 
$(x_0,y_0)\in X$ and
$rank A(x_0)-y_0 I_r \leq r-2$. If we denote by
$\Delta _{ij}(x,y)$ the $(i,j)$th minor of the matrix $A(x)-yI_r$, then we have
$$\Delta _{ij}(x_0,y_0)=0, $$
$$
P'_x(x_0,y_0)= \sum_{i,j} (-1)^{i+j}a_{ij}'(x)\Delta _{ij}(x_0,y_0)=0,
$$
$$
P'_y(x_0,y_0)= -\sum_{i} \Delta _{ii}(x_0,y_0)=0
$$
and hence the curve $X$ is not smooth at $(x_0,y_0)$. 
The regularity of $A_d$ is proved in the same way, we only 
change the local 
coordinates on the weighted projective space $\bbbp^2(d)$ as
$$
x\rightarrow \frac{1}{x}, y\rightarrow \frac{y}{x^d} \; .
$$

Let $A(x)$ be a matrix with a spectral polynomial $P(x,y)$. By 
Proposition \ref{regular} the leading term
$J$ of $A(x)$ is a regular matrix. The characteristic polynomial of $J$ 
obviously coincides with the highest order weight-homogeneous part of  $P(x,y)$.
If $m= \sum n_i p_i$ is the modulus of the spectral curve, then without loss of
generality we shall suppose that
 $J= diag(J(\lambda_1),J(\lambda_2),...,J(\lambda_s))$, where
$J(\lambda_i)$ is a Jordan block of dimension $n_i$ with eigenvalue $\lambda_i$, 
and $\lambda_i\neq \lambda_j$. 

We recall first the definition of the eigenvector map $l$.
Consider a line bundle over a curve $X$ which is a sub-bundle of the trivial vector bundle
$X\times \bbbc^r$. It is defined by a meromorphic vector $f(p)=(f_1,f_2,...,f_r)$.
We shall always suppose that $f$ is {\it normalized}, that is to say the meromorphic functions
$f_1,f_2,...,f_r$ have not a common zero.The map
$$
X \rightarrow \bbbp^{r-1} : p \rightarrow [f_1(p),f_2(p),...,f_r(p)]
$$
is holomorphic, so we obtain a holomorphic line bundle over $X$. 
We denote its dual by $L$.
If $D$ is the pole divisor of $f$, that is to say the minimal effective
divisor such that $(f_i) \geq -D$ for any $i$, then $L=L_D$.

\begin{definition} (eigenvector line bundle  on the  spectral curve
$X$)

Let $f(x,y)=^t(f_1,f_2,...,f_r)$, $p=(x,y) \in X$ be a normalized
 eigenvector of the matrix $A(x)\in M_P$, $A(x)f=yf$. It defines a
 line bundle over the 
spectral curve
$X$ which will be  called eigenvector line bundle.
 Denote its dual by
$L$ and  the corresponding sheaf of sections by $\bbbL$. 
\end{definition}

Of course if $D$ is the pole divisor
of the normalized eigenvector $f$ then $L=L_D$ and $\bbbL = \bbbL(D)$.
The following properties of $L$ are well known (see \cite{Beauville,Adler,Reyman})
\begin{pr}
\label{ff}
If $\pi: X \rightarrow \bbbp^1$ is the projection defined above, then  the sheaf $\pi_*\bbbL$ is a trivial
$\bbbO_{\bbbp^1}$ module of rank $r$
$$
\pi_*\bbbL= \bbbO_{\bbbp^1}\oplus\bbbO_{\bbbp^1}\oplus...\oplus \bbbO_{\bbbp^1}
$$
and the functions $f_1,f_2,...,f_r$ form a basis of global sections of 
$\pi_*\bbbL$. Moreover  $deg(L)=deg(D)= p_a=p_g+r-1$, 
where $p_g= (r-1)(dr-2)/2$, 
$dim H^0(X,\bbbL) = r$ and $f_1,f_2,...,f_r$ form a basis of $H^0(X,\bbbL(D))$.
\end{pr}

The above proposition shows that if the matrices $A(x)$, $\tilde{A}(x)$ define isomorphic
eigenvector bundles, then for some $R\in {\bf GL}_r(\bbbc)$, 
$A(x)=R\tilde{A}^(x)R^{-1}$. Indeed, if $f$, $\tilde{f}$ are the corresponding
 normalized
eigenvectors with equivalent pole divisors  $D\sim\tilde{D}$, then
there
exists 
a meromorphic function
$\varphi$
on $X$ such that $(\varphi)=\tilde{D}-D$. As
 $\varphi \tilde{f_i}$ form a basis
of $H^0(X,\bbbL(D))$, then there exists a matrix $R\in {\bf GL}_r(\bbbc)$ such that 
$f=\varphi R \tilde{f}$, and hence $A(x)=R\tilde{A}^(x)R^{-1}$. Thus we obtain
a holomorphic map

\begin{eqnarray*}
&\{ \mbox{ a matrix } A(x)\in M_P \mbox{ up to conjugation by a matrix in } 
\bbbp{\bf GL}_r(\bbbc) \} & \\
 &\downarrow l & \\
 &\{\mbox{ an isomorphism class of a line bundle } L \in Pic^{p_a}(X)-\Theta\} .&
\end{eqnarray*}
The following beautiful argument of Beauville \cite{Beauville} shows that $l$
is a bijection
(see also section 8.2. of the survey \cite{Reyman} by A.G.Reyman and 
M.A. Semenov-Tian-Shansky). Take a degree $p_a=p_g+r-1$ invertible sheaf 
 $\bbbL$  on $X$. By Riemann-Roch theorem 
$$
\chi(\bbbL)= deg(\bbbL)-p_g +1=r ,
$$
$$
\chi(\pi_*\bbbL)=deg(\pi_*\bbbL) + (1-p_g(\bbbp)) rank(\pi_*\bbbL)=
deg(\pi_*\bbbL) +r ,
$$
 by Grothendieck-Riemann-Roch
$$
\chi(\bbbL)= \chi(\pi_*\bbbL),
$$
and hence
$
deg(\pi_*\bbbL)=0
$.
If we suppose in addition that $\bbbL \in Pic^{p_a}(X)-\Theta$ we conclude that
$\pi_*\bbbL$ is the rank $r$  trivial vector bundle $(\bbbO_{\bbbp^1})^r$.
The invertible sheaf $\bbbL$ on $X$ can be equivalently described as a locally
 trivial
$\bbbO_{\bbbp^1}$ module $\pi_* \bbbL$ equipped with an additional structure of a 
$\pi_* \bbbO$
module, that is to say a homomorphism of algebras 
$a: \pi_* \bbbO \rightarrow {\it End }(\pi_* \bbbL)$. 
To describe the homomorphism $a$ amounts to give a linear map (multiplication by $y$)
$$
 \pi_* \bbbL \rightarrow \pi_* \bbbL(d)
$$
that is to say a polynomial $r\times r$ matrix  of degree $d$. Denote the transposed
to this matrix by $A(x)$.
Clearly $A(x)$ satisfies
$P(x,A(x))=0$ and as $P(x,y)$ is irreducible over $\bbbc(x)$ then by the Cayley-Hamilton theorem the spectral
polynomial of $A(x)$ is $P(x,y)$. 
Note that the matrix $A(x)$ is determined only modulo an automorphism
of $\pi_* \bbbL$. Thus the matrix $A(x)$ is
determined only up to conjugation $A(x) \rightarrow R^{-1}A(x)R$ by a matrix 
$R \in \bbbp{\bf GL}_r(\bbbc)$.

The next step is to define the eigenvector line bundle on the singularized spectral
curve $X'$ and the corresponding map $l'$.

\begin{definition} (eigenvector sheaf  on the singularized spectral curve
$X'$)
\label{egv}
Let $f(x,y)$ be an eigenvector of $A(x)$ normalized by the condition
$$
\sum_{i=1}^r f_i \equiv 1
$$
and let $D$ be the minimal divisor, such that $(f_i)\geq -D$, $i=1,2,...s$. 
Then $D$ 
is an effective divisor, $D \subset X_{reg}$, and we define  the  invertible  
eigenvector sheaf
 on the singularized spectral curve $X'$ to be $\bbbL'= \bbbL'(D)$, where
$$
\bbbL'_p(D) = 
\left\{
\begin{array}{c}
\bbbL_p(D), \; \; p \neq \infty \\
\bbbO'_p, \; \; p = \infty \; . 
\end{array} 
\right.
$$
\end{definition}

We denote by $L'$ the line bundle over $X'$ associated to the invertible sheaf
$\bbbL'$. To prove the correctness of the above definition it remains to check that
$D \subset X_{reg}$, where $D$ is the pole divisor of the normalized eigenvector
$f$.

 Let $S = \sum_{1}^s p_i$ be the support of 
the modulus
$m=\sum_{1}^s n_ip_i$ and we may suppose that $p_i \in X$ corresponds to the Jordan block
$J(\lambda_i)$ of the matrix $J$. An easy computation shows that $f(p_i)$ 
determines a line over
$p_i$, collinear with the eigenvector $(0,0,..,0,1,0,...,0)$ of $J$ corresponding to the Jordan block
$J(\lambda_i)$. The eigenvector $f$ has a pole at $p_i$ if and only if the line determined by $f(p_i)$
is contained in the plane $f_1+f_2+...+f_r=0$, so $p_i$ is not a pole.

Let $Pic^{p_a}(X')$ be the ``shifted" Picard group $Pic^0(X')=J(X')$ of 
degree $p_a$ line bundles on $X'$.
It is isomorphic to the Jacobian variety $J(X')$ 
and  $J(X')-\Theta'$ is the subset of line bundles
$L' \in Pic^{p_a}(X')$ with one non-zero holomorphic section 
$h^0\bbbL'=dim H^0(X',L')=1$. Definition \ref{egv} establishes a 
holomorphic map
\begin{eqnarray*}
&\{ \mbox{ a matrix } A(x)\in M_P^J \}
 &\\
& \downarrow l' & \\
 &\{ \mbox{ an isomorphism class of a line bundle }
L' \in Pic^{p_a}(X')-\Theta'\} .&
\end{eqnarray*}
Clearly the map $l'$ is such that the diagram (\ref{commute}) commutes:
$\phi\circ l'= l \circ \varphi$. 
As the map $l$ is a bijection, then
to show that $l'$ is a bijection too it suffices to check
that

i) the fibres $\varphi^{-1}(b)$ and $\phi^{-1}\circ l(b)$ have the same dimension.

ii) $l': \varphi^{-1}(b) \rightarrow \phi^{-1}\circ l(b)$ is an injective
homomorphism of algebraic groups.

Step i) is obvious and the dimension of the fibres is $r$. To check that
$l'$ is injective we take a sheaf
$\bbbL' \in Pic^{p_a}(X')-\Theta'$ in the image of $l'$.
 By Riemann-Roch theorem \cite{Serre}
$$
\chi(\bbbL')= deg(\bbbL')-p_a +1=1 ,
$$

$$
 \chi(\pi_*\bbbL')= deg(\pi_*\bbbL') + (1-p_g(\bbbp)) rank(\pi_*\bbbL')=
deg(\pi_*\bbbL') +r ,
$$
by Grothendieck-Riemann-Roch 
$$\chi(\bbbL')=\chi(\pi_*\bbbL').$$
 We conclude
that
 $\pi_*\bbbL'$ is 
a degree
$1-r$ and rank $r$
locally trivial $\bbbO_{\bbbp^1}$ module,  having one holomorphic section, 
 $h^0\pi_* \bbbL' =h^0 \bbbL'=1$, so
$$
\pi_* \bbbL' = \bbbO_{\bbbp^1}\oplus\bbbO_{\bbbp^1}(-1) \oplus...\oplus\bbbO_{\bbbp^1}(-1) \; .
$$
The invertible sheaf $\bbbL'$ on $X'$ can be equivalently described as a locally trivial
$\bbbO_{\bbbp^1}$ module $\pi_* \bbbL'$ equipped with an additional structure of a $\pi_* \bbbO'$
module, that is to say a homomorphism of algebras 
$a: \pi_* \bbbO' \rightarrow {\it End }(\pi_* \bbbL')$. It is easy to compute $\pi_* \bbbO'$: a basis
over the affine plane $\bbbc$ is given by $\{ 1,y,y^2,...,y^{r-1}\}$ and over $\bbbp-\{0\}$ by
$\{ 1,y/x^{d+1},y^2/x^{2d+1},...,y^{r-1}/x^{(r-1)d+1} \}$, so
$$
\pi_* \bbbO'= \bbbO_{\bbbp^1}\oplus \bbbO_{\bbbp^1}(-d-1)\oplus...\oplus\bbbO_{\bbbp^1}(-d(r-1)-1) \; .
$$
To describe the homomorphism $a$ amounts to give a linear map (multiplication by $y$)
$$
 \pi_* \bbbL' \rightarrow \pi_* \bbbL'(d)
$$
that is to say a polynomial $r\times r$ matrix  of degree $d$. 
Denote  the transposed to this matrix by $A(x)$. If $f_1,f_2,...f_r$ is a
normalized basis of $\pi_* \bbbL'$ over $\bbbc$, $\sum f_i \equiv 1$,
then $^t(f_1,f_2,...f_r)$ is an eigenvector of $A(x)$.
Clearly $A(x)$ satisfies
$P(x,A(x))=0$ and as $P(x,y)$ is irreducible over $\bbbc(x)$ then by the Cayley-Hamilton theorem the spectral
polynomial of $A(x)$ is $P(x,y)$. 
The homomorphism $a$ is determined modulo an automorphism
of $\pi_* \bbbL'$. 
In the base $f_1,f_2,...,f_r$ 
the vector $1 \equiv \sum f_i \in H^0(\bbbp, \pi_* \bbbL')$
has coordinates $e= ^t(1,1,...,1)$, and hence the group $Aut(\pi_* \bbbL')$ is
identified to 
$$
{\bf GL}_r(\bbbc;e)= \{ R\in {\bf GL}_r(\bbbc) : e
\mbox{ is an eigenvector of } R \} .
$$
As we supposed that $\pi_* \bbbL' \in l'(M_P^J)$, then in a suitable basis of
$\pi_* \bbbL'$ we have $A(x)\in M_P^J$.
If $\tilde{A}(x)\in M_P^J$ is another matrix
 which defines the same eigenvector sheaf $\bbbL'$, then
$\tilde{A}(x)=R A(x) R^{-1}$ for some $R\in \bbbp{\bf GL}_r(\bbbc;e)$.
As at the same time $R\in \bbbp{\bf GL}_r(\bbbc;J)$, we conclude that
 $R= 1\in \bbbp{\bf GL}_r(\bbbc)$.

Finally we note that the vector fields $[J^k,A(x)]$, $r=1,...,r-1$ are tangent
 to the fibre
$\varphi^{-1}(b)$, $\bbbp{\bf GL}_r(\bbbc;J)$ invariant and linearly
independent (this follows from the regularity of $J$). The images of these 
vector fields in $Pic^{p_a}(X') \sim J(X')$ are well known to be translation
invariant \cite{Reyman} and hence
 $l': \varphi^{-1}(b) \rightarrow \phi^{-1}\circ l(b)$ is a homomorhism.
This completes the proof of step ii).

It remains to prove that $M_P^J $ is a smooth manifold, 
that is to say, to find at any point $A(x)\in M_P^J$
vector fields which span the tangent space, and such that their images in 
$J(X')-\Theta'$ span the tangent space too. These vector fields are given 
by
$$
Y^{(i)}_a(A(x))= [\frac{A^{i}(a)}{x-a},A(x)], \; a\in \bbbp, \; i\in \bbbn 
$$
but this will be explained in  the next section.$ \Box$

\section{ Integrable Systems }
\label{integrable}

Let us fix a non-zero matrix $J \in {\bf gl}(r,\bbbc)$ and denote by
$M^J$ the affine space of all matrix polynomials $A(x)$ of the form
$$
A(x)= J x^d+A_{d-1}x^{d-1}+...+A_0 , \; A_i \in {\bf gl}_r(\bbbc) 
\; .$$
The space $M^J$ is of dimension $dim M^J= 2p_a+ dr = dr^2$ and it
 carries several compatible Poisson structures of rank
$2p_a= dr(r-1)$. Let us fix such a structure $\{.,.\}$. A function $\varphi$ on
$M^J$ is called invariant if it is constant on each isospectral manifold
$$ M_P^J= \{A(x)\in M^J: det(A(x)-y I_r)= P(x,y) \} .$$
The algebra of invariant functions on $M^J$ is thus generated by the
$dr(r+1)/2$ non-trivial coefficients of $P(x,y)$ (which are in addition functionally
independent).

It turns out that the invariant functions commute with respect to $\{.,.\}$.
Moreover, the tangent space to $M^J_P$ at any point $A(x) \in M^J_P$ is the span
of all Hamiltonian vector fields $X_\varphi= \{.,\varphi \}$, where $\varphi$ is
an invariant function. It follows that any such Hamiltonian vector field
 $X_\varphi$ is completely integrable in the sense of Liouville, 
and hence
its solutions can be explicitly computed ``by quadratures".

The purpose of this section is to describe 
briefly the Hamiltonian structure of the completely integrable system (\ref{axa})
 (thus 
justifying the title of the article). The scheme is quite classical now and proofs
together with historical comments may be found in \cite{Reyman}.

 We describe first the compatible Poisson structures. 
Let
$$\tilde{{\bf g}}= {\bf g}[x,x^{-1}]$$
be the loop algebra of the Lie algebra ${\bf g}$  formed by  Laurent polynomials
in one variable $x$ with coefficients in ${\bf g}$, and commutator
 given by
$$
[\sum_iA_ix^{i},\sum_jB_j x^j]=\sum_k (\sum_{i+j=k}[A_i,B_j])x^{k}, A_i,B_j \in {\bf g} \; .
$$
Let $$\tilde{{\bf g}}^* = {\bf g}^* [x,x^{-1}]$$
be the ``restricted" dual space to $\tilde{{\bf g}}$ consisting of
Laurent polynomials.The space $\tilde{{\bf g}}^*$ carries a canonical
Lie-Poisson structure, which is the extension of the Lie algebra of linear
functions on $\tilde{{\bf g}}^*$ to the entire space of smooth functions on
$\tilde{{\bf g}}^*$ (a linear function on $\tilde{{\bf g}}^*$
is identified to a point in $\tilde{{\bf g}}$). 
Any non-degenerate
invariant bi-linear form $<.,.>$ on $\tilde{{\bf g}}$
identifies $\tilde{{\bf g}}^*$ to $\tilde{{\bf g}}$ so the 
latter space also carries a Poisson structure. To be explicit, put
${\bf g}=\tilde{{\bf gl}}_r(\bbbc)$ and
$$<A(x),B(x)> = Res_{x=0} Trace(A(x)B(x)) \frac{dx}{x} .$$
Choose a basis
$e^{a}$ in ${\bf g}$ and let $C^{ab}_c$ be the structure constants of
${\bf g}$, $[e^{a},e^b]= \sum_c C^{ab}_c e^c$. Let 
$A(x)=\sum_i A_i x^{i} \in \tilde{{\bf g}}$, where
$A_i= \sum_a A_i^{a}e^{a}$. Then
$$
\{A_i^{a},A_j^{b}\}= -\sum_c C^{ab}_c A^c_{i+j} \; .
$$
The simplectic leaves of this Poisson structure are the co-adjoint orbits of 
the Lie group underlying  $\tilde{{\bf g}}$. The corresponding ring of adjoint
invariants (Casimir functions) is generated by
$$
\varphi_{mn}(A(x))= Res_{x=0}(x^{-n}\varphi(x^mA(x)))dx, \; m,n\in \bbbn 
$$
where $\varphi$ is any invariant function on ${\bf g}$.
It is clear that in such a way any Lie algebra structure on $\tilde{{\bf g}}$ 
defines  a Poisson structure on $\tilde{{\bf g}}$. The most important class
of Poisson brackets are the so called $R$-brackets. Namely, let 
$R\in End(\tilde{{\bf g}})$ be a linear operator, and suppose that the commutator
\begin{equation}
\label{xy}
[X,Y]_R= \frac{1}{2}([RX,Y]+[X,RY]) , X,Y \in \tilde{{\bf g}}
\end{equation}
satisfies the Jacobi identity (this happens for example if $R$ satisfies the 
classical Yang-Baxter identity). This induces, according to the scheme described above,
 a Poisson bracket $\{.,.\}_R$
on $\tilde{{\bf g}}$. The importance of the $R$-bracket is related to the
following result (due to Semenov-Tian-Shanski \cite{Semenov,Reyman}
and closely related to the so called Adler-Kostant-Symes theorem).
\begin{th1}\mbox{}
 \\(i) The Casimir functions $\varphi_{mn}$ on $\tilde{{\bf g}}$ are in involution
with respect to the $R$-bracket.\\
(ii) The Hamiltonian system associated to 
$H(A(x))= Res_{x=0} Trace( \frac{A^{k+1}(x)}{k+1}) \frac{dx}{x}$,  $\{.,.\}_R$,  is
given by
\begin{equation}
\label{llax}
\frac{d}{dt} A(x)= [A(x), M], 
A(x)\in \tilde{{\bf g}}, M= \frac{1}{2}R( A^k(x)) \; .
\end{equation}
\end{th1}
The decomposition
$$\tilde{{\bf g}} = \tilde{{\bf g}}^+ \oplus\tilde{{\bf g}}^-$$
where
$$
\tilde{{\bf g}}^+= \oplus_{i=0}^{\infty} \tilde{{\bf g}} x^{i}, \;
\tilde{{\bf g}}^-= \oplus_{i=-1}^{-\infty} 
\tilde{{\bf g}} x^{i} $$
defines a $R$ matrix on ${\bf g}$. Namely,
if $A(x)=A(x)^++A(x)^- \in \tilde{{\bf g}}$ where
$A(x)^\pm \in \tilde{{\bf g}}^\pm$, then define
$$R(A(x))=A(x)^+-A(x)^- \; .$$
The commutator (\ref{xy}) is given by
$$
[A(x),B(x)]_R =[A(x)^+,B(x)^+]-[A(x)^-,B(x)^-]
$$
and it
satisfies the Jacobi identity. The induced Poisson bracket $\{.,.\}_R$ on 
$\tilde{{\bf g}}$ is explicitly given (in the notations above) by
$$
\{A_i^{a},A_j^{b}\}= - \epsilon_{ij}\sum_c C^{ab}_c A^c_{i+j} 
$$
where $\epsilon_{ij}=1$ for $i,j\geq 0$, $\epsilon_{ij}=-1$ for $i,j< 0$, and
$\epsilon_{ij}=0$ for $i\geq 0, j<0$. 

Let $$q(x)= \sum_{i=-d+1}^1 q_ix^{i}$$ be a fixed polynomial. Then the embedding
\begin{equation}
\label{emb}
M^J \hookrightarrow  \tilde{{\bf g}} : A(x) \mapsto q(x)A(x)
\end{equation}
is a {\it Poisson mapping} with respect to the $R$-bracket. This means that the 
embedding induces a Poisson structure on $M^J$. We obtain thus a family of
{\it compatible} Poisson structures on $M^J$ which depend linearly on the coefficients
$q_i$ of the polynomial $q(x)$.
 
\begin{cor}
The Hamiltonian system on $M^J$ associated to 
$$H(A(x))= Res_{x=0} Trace(\frac{A^{k+1}(x)}{k+1}) \frac{dx}{x}$$
 and to the Poisson
structure induced by the embedding (\ref{emb}) is given by
\begin{equation}
\label{corl}
\frac{d}{dt} A(x)= [A(x), M_\pm], 
A(x)\in \tilde{{\bf g}}, M_\pm= \pm(q(x) A^k(x))_\pm \in \tilde{{\bf g}}^\pm \; .
\end{equation}
\end{cor}
If we choose for example $q(x)=1/x$, then (\ref{corl}) takes the form
\begin{equation}
\label{corl1}
\frac{d}{dt} A(x)= [\frac{A^k(0)}{x},A(x)]   .
\end{equation}
The construction of the loop algebra $\tilde{{\bf g}}$ was related to the choice
of Laurent polynomials with a pole at $x=0$. It is obvious that all that holds true
if we consider Laurent polynomials with a pole at $x=a\in \bbbc$. In this case
(\ref{corl1}) takes the Beauville form
\begin{equation}
\label{corl2}
\frac{d}{dt} A(x)= [\frac{A^k(a)}{x-a},A(x)]= Y^{(k)}_a(A(x)) .
\end{equation}
Recall now that when the spectral curve is smooth, then the invariant level set
(the isospectral manifold of $A(x)$) of (\ref{corl2}) is smooth and bi-holomorphic
to the Zariski open subset $J(X')-\Theta'$ of
the generalized Jacobian $J(X')$ (Theorem \ref{gav}). It is shown in \cite{Reyman}
that the vector fields $Y^{(k)}_a(A(x))$ induce  translation invariant vector
fields on $J(X')$ (although the results are formulated only on $J(X)$). The direction
of $Y^{(k)}_a(A(x)) $ is moreover explicitly computed (formula (8.5) on p.177,
but see also \cite[Corollary 2.7]{Beauville}). These formulae imply that the
vector fields  $Y^{(k)}_a(A(x))$ span, for generic $a$ the tangent space to the generalized
Jacobian $J(X')$. 

We conclude that the Hamiltonian system (\ref{corl}) is completely integrable.
\begin{definition}
A Hamiltonian system is called algebraically completely integrable, provided that
it is completely integrable, and in addition each generic complex invariant level set
is a Zariski open subset of a commutative algebraic group, on which the Hamiltonian
vector fields generated by the first integrals are translation invariant.
\end{definition}
Of course in order that the above definition makes a sense we must suppose that
the Poisson manifold, the Hamiltonian functions and vector fields are 
{\it algebraic} (compare to \cite[p.3.53]{Mumford}).
Taking into account the results of section \ref{general} we obtain
\begin{cor}
\label{corint}
The  system (\ref{corl}) (and hence (\ref{axa}) ) is a $p_a=dr(r-1)/2$ degrees of freedom
algebraically completely integrable Hamiltonian system. 
\end{cor}

\section{Examples}
\label{Examples}

Let $X$ be a smooth elliptic curve, $m=P_1+P_2$, $P_1\neq P_2$, an effective divisor
 on 
$X$, and $X'$ the corresponding singularized curve. 
The generalized Jacobian $J(X')$ is an extension of the usual Jacobian $J(X)$
by $\bbbc^*$
$$
0 \rightarrow  \bbbc^*  \rightarrow  J(X')  \rightarrow  J(X) \rightarrow  0
$$
and it is easy to check that
the above extension
is {\it never} trivial. Indeed, if the generalized Jacobian 
$J(X')$ is isomorphic to $J(X) \times \bbbc^*$ then 
$J(X') = \bbbc^2 / \Lambda$ where
$$
\Lambda= \bbbz \left( \begin{array}{c}
2 \pi  \\
0
\end{array} 
\right)
+
\bbbz \left( \begin{array}{c}
 0 \\
2\pi
\end{array} 
\right)
+
\bbbz \left( \begin{array}{c}
 \tau_1 \\
 \tau_2
\end{array} 
\right)
$$
with $\tau_2=0$. 
The generalized Riemann theta
function \cite{Belokolos,Fedorov}
$$
\tilde{\theta}(z_1,z_2 | \tau_1,\tau_2) = 
e^{z_2/2}\theta(z_1+\tau_2/2 | \tau_1)
+e^{-z_2/2}\theta(z_1-\tau_2/2 | \tau_1), (z_1,z_2)\in \bbbc^2 / \Lambda
$$
decomposes into the product 
$(e^{z_2/2}- e^{z_2/2})\theta(z_1| \tau_1)$
where
$\theta(z_1 | \tau_1)$ is the
usual elliptic Riemann theta function ( $z_1\in \bbbc/\{2 \pi i \bbbz+ \bbbz \tau_1 \}$). It follows that the generalized Riemann
theta divisor $(\tilde{\theta})$ is reducible which contradicts to the fact that it 
is isomorphic to the affine curve $X-\{P_1 \cup P_2\}$ \cite{Fedorov}.

Consider now the affine space $M^J$ of matrix polynomials $A(x)$ of the form
$$
A(x)=J x^2 + A_1 x+A_0, \; A_0,A_1 \in {\bf gl}_2(\bbbc) \;
$$
where $J$ is a fixed matrix with distinct eigenvalues.
As we explained in section \ref{integrable} the Lax pair
\begin{equation}
\label{lgr}
\frac{d}{dt} A(x)= [A(x),\frac{A(a)}{x-a}]=[A(x),Jx+aJ+A_1]=
[A(x),((q(x)A(x))_+], \; q(x)=\frac{x+a}{x^2}
\end{equation}
defines  a completely integrable Hamiltonian system on the simplectic leaves of several
compatible Poisson structures on $M^J$. Moreover, 
when the spectral curve $X$ with affine equation $\{(x,y)\in \bbbc^2:P(x,y)=0\}$
 is 
smooth,
the corresponding isospectral manifold 
$M_P^J = \{A(x) \in M^J: det(A(x)-yI_2)=P(x,y) \}$
is smooth and is described as in 
Theorem \ref{gav}. In addition the above vector field is translation invariant on
the generalized Jacobian $J(X')$, so our system is algebraically completely 
integrable. As the modulus of the spectral curve $X$ is
$m=\infty^+ + \infty^-$, 
where $\infty^\pm$
are the two ``infinite" points on $X$, then the
generalized
Jacobian $J(X')$ is described as above.
 
Our purpose is to show that, for appropriate
choice of the  matrix $J$ and the parameter $a$,
equation (\ref{lgr})
 is the classical
equation of heavy symmetric top.  The symmetry group $\bbbc^*$ is then the 
complexified
circle action (rotations about the symmetry axe of the top).  
 In the sequel we put
 $$J=\frac{\sqrt{2}}{\epsilon} \left(
\begin{array}{lr}
0& 1 \\
1 &0
\end{array} 
\right), \epsilon = \exp{\sqrt{-1}\pi/4}  $$
Consider the isospectral manifold
$$
M_f^J = \{ A(x) \in M^J : det(A(x)- yI_2) = y^2-f(x) \}
$$ 
where $f(x)$ is a fixed monic polynomial
$$
f(x)=x^4+a_1x^3+a_2x^2+a_3x+a_4 \; 
$$
without double roots. We may consider $M_f^J$ as a
subvariety of the affine vector space of traceless matrices
$$
V= \{A(x) \in M^J: Trace(A(x))=0 \} \; .
$$
By making use of the isomorphism of Lie algebras $sl_2(\bbbc)$ and $so_3(\bbbc)$
given by
$$
\left(
\begin{array}{ccc}
0 & -z& y \\
z& 0& -x\\
-y & x & 0
\end{array} \right)
\rightarrow
\frac{1}{\sqrt{2}}
\left(
\begin{array}{cc}
\epsilon x&\epsilon z + \overline{\epsilon} y  \\
\epsilon z - \overline{\epsilon} y & -\epsilon x
\end{array} \right) ,\;
\epsilon = \exp{\sqrt{-1}\pi/4} 
$$
we may identify $V$ to the affine space
$$
\{ L(x): L(x)= \chi x^2+ Mx-\Gamma, M,\Gamma \in so_3(\bbbc),
\chi=\left(
\begin{array}{ccc}
0 & -1& 0 \\
1& 0&0\\
0 & 0 & 0
\end{array} \right) \}  .
$$
The Hamiltonian system (\ref{lgr}) takes the form
\begin{equation}
\label{LaxL}
\frac{d}{dt}(x^2 \chi +x M -\Gamma)=
[x^2 \chi+x M -\Gamma, x \chi+ M + aJ ] \; .
\end{equation}
If we put at last 
$a=-m \Omega_3$ and
$$
 M=(\Omega _1,\Omega _2,(1+m)\Omega _3),
\Gamma=(\Gamma_1,\Gamma_2,\Gamma_3), \Omega =(\Omega _1,\Omega _2,\Omega _3)
$$
then we obtain
$$
\frac{d}{dt} M= [M,\Omega]-[\Gamma,\chi], \frac{d}{dt} \Gamma = [\Gamma,\Omega].
$$
which are the equations describing the Lagrange top. Indeed, after identifying the
isomorphic 
Lie algebras  $(\bbbr^3, \wedge)$ and $(so(3),[.,.])$, and  making obvious rescalings
we obtain the system
\begin{equation}
\label{EP}
\frac{d}{dt} M =     M\times\Omega + \chi\times\Gamma , \; \;
\frac{d}{dt} \Gamma =\Gamma\times\Omega \; 
\end{equation}
where
$$M=(I_1\Omega _1,I_2\Omega _2,I_3\Omega _3),\Omega=(\Omega _1,\Omega _2,\Omega _3), 
\Gamma=(\Gamma_1,\Gamma_2,\Gamma_3), \chi=(\chi_1,\chi_2,\chi_3)\; $$
and in addition
$
I_1=I_2, \chi_1=\chi_2=0$. 
Here $M,\Omega $ and $\Gamma$ denote respectively the angular momentum, the angular velocity and
 the coordinates of the unit vector in the 
direction of gravity, all expressed in body-coordinates. 
The constant vector $\chi$ is the center 
of mass in body-coordinates multiplied by the mass  and the 
acceleration, $I_1,I_2,I_3$ are the principal moments of inertia of the body.

To resume, we proved that the Lagrange top is an algebraically completely integrable
system. It linearizes on a two-dimensional complex algebraic group --- 
the generalized Jacobian 
$J(X')$ of 
an elliptic curve $X$ with two points $\infty^\pm$  identified.
This result is proved directly in \cite{Gav}.
If we reduce further the system (\ref{EP}) with respect to the circle action
$\bbbc^*$ we obtain, as it is well known, a one degree of freedom algebraically completely integrable system
linearized on the elliptic curve $J(X')/\bbbc^* = X$ \cite{Lagrange,Adler,Ratiu,Verdier,Audin2}.
Other mechanical systems linearized on non-compact algebraic groups were recently
studied by Fedorov \cite{Fedorov}. 

\paragraph
{The general integral of a system of hyperelliptic differential equations}

Let $f(x)$ be a fixed polynomial of degree $2n$ or $2n-1$
without double roots and consider the following 
{\it hyperelliptic system of differential equations}
\begin{eqnarray}
\label{hyper}
\frac{dx_1}{\sqrt{f(x_1)}}+\frac{dx_2}{\sqrt{f(x_2)}}+...+
\frac{dx_n}{\sqrt{f(x_n)}}&=&0 \nonumber \\
\frac{x_1dx_1}{\sqrt{f(x_1)}}+\frac{x_2dx_2}{\sqrt{f(x_2)}}+...+
\frac{x_ndx_n}{\sqrt{f(x_n)}}&=&0  \\ 
\ldots & & \ldots \nonumber\\
\frac{x_1^{n-2}dx_1}{\sqrt{f(x_1)}}+\frac{x_2^{n-2}dx_2}{\sqrt{f(x_2)}}+...+
\frac{x_n^{n-2}dx_n}{\sqrt{f(x_n)}} &=&0 \nonumber
\end{eqnarray} 
Suppose that the polynomial $f(x)$ is written in the form
$$f(x)=-A^2(x)+B^2(x)+C^2(x)$$ where 
\begin{equation}
\label{axb}
A(x)= \sum_{k=0}^na_kx^k,\;B(x)= \sum_{k=0}^nb_kx^k,\;
C(x)= \sum_{k=0}^nc_kx^k \; .
\end{equation}
Jacobi \cite{Jacobi} proved in 1846 the following
\begin{th1}
\label{Jacobi}
Let $x_1(\varphi),x_2(\varphi),...,x_n(\varphi)$ be the roots of the polynomial
equation
$$
A(x)=B(x) cos(\varphi)+C(x) sin(\varphi) \; .
$$
Then ${\bf x}(\varphi)=(x_1(\varphi),x_2(\varphi),...,x_n(\varphi) )$ is an
integral curve of (\ref{hyper}).
\end{th1}

Let us note that the phase space of the system (\ref{hyper}) is the $n$th
symmetric product $S^n\Gamma $ of the smooth affine curve
$$
\Gamma = \{(x,y): y^2=f(x)\} .
$$
The variables $x_1,x_2,...,x_n$ provide a system of local coordinates in a 
neighborhood of any generic point on the smooth manifold $S^n\Gamma $.
We shall give an independent proof of Jacobi's theorem in the light of the
 present paper. For a further discussion on the Jacobi's paper 
see Mumford \cite[p. 3.17]{Mumford}.

Assume first that $deg(f)=2n$ and
consider, instead of system (\ref{hyper}), the following generalized Jacobi
inversion problem \cite{Clebsch,Krazer,Serre}

\begin{eqnarray}
\label{hyper1}
\frac{dx_1}{\sqrt{f(x_1)}}+\frac{dx_2}{\sqrt{f(x_2)}}+...+
\frac{dx_n}{\sqrt{f(x_n)}}&=&dz_1 \nonumber \\
\frac{x_1dx_1}{\sqrt{f(x_1)}}+\frac{x_2dx_2}{\sqrt{f(x_2)}}+...+
\frac{x_ndx_n}{\sqrt{f(x_n)}}&=&dz_2  \\ 
\ldots & & \ldots \nonumber\\
\frac{x_1^{n-1}dx_1}{\sqrt{f(x_1)}}+\frac{x_2^{n-1}dx_2}{\sqrt{f(x_2)}}+...+
\frac{x_n^{n-1}dx_n}{\sqrt{f(x_n)}} &=&dz_n \nonumber
\end{eqnarray} 
 It involves the differential of second kind 
\begin{equation}
\label{diff}
\frac{x^{n-1}dx}{\sqrt{f(x)}}
\end{equation}
 on the completed and normalized genus
$n-1$ hyperelliptic curve $X= \bar{\Gamma }$
$\Gamma =\{(x,y): y^2=f(x)\}.$ Put $m=\infty^++\infty^-$,
where $\infty^\pm$ are the two poles of the differential (\ref{diff}) and let $X'$ be the singularized
curve $X$ relative to the effective divisor $m$.
 The general symmetric function in 
$x_i, \sqrt{f(x_j)}$ can be expressed as a meromorphic function in
$^t(z_1,z_2,...,z_n) \in J(X')=\bbbc^n/\Lambda$, where $\Lambda$ is the $\bbbz$ lattice
$$
\Lambda= \{^t( \oint_\gamma \frac{dx}{\sqrt{f(x)}},
\oint_\gamma \frac{xdx}{\sqrt{f(x)}},...,
\oint_\gamma \frac{x^{n-1}dx}{\sqrt{f(x)}}) \}_\gamma, \;
\gamma \in H_1(X-\{\infty^+,\infty^-\},\bbbz\} \; .
$$
The generalized Jacobian $J(X')$ is  
a $\bbbc^*$ extension of the usual Jacobian $J(X)$
\begin{equation}
\label{hextension}
0\stackrel{exp}{\rightarrow} \bbbc^* 
 \stackrel{ }{\rightarrow}
J(X')\stackrel{\phi}{\rightarrow}
J(X) \rightarrow 0 
\end{equation}
where $\phi$ is the projection $\phi(z_1,z_2,...,z_n)= ^t(z_1,z_2,...,z_{n-1})$
(see section \ref{general}).

It follows that {\it an integral curve of the system (\ref{hyper}) is just the fibre
$\phi^{-1}(z_1^0,z_2^0,...,z_{n-1}^0) $ over the point
$^t(z_1^0,z_2^0,...,z_{n-1}^0) \in J(X)$. In particular each integral curve is 
isomorphic to the algebraic group $\bbbc^*$, and the set of all integral curves is 
parameterized by the  Jacobian variety $J(X)$. }

Theorem \ref{gav} provides an explicit parameterization of the fibre
$\phi^{-1}(z_1^0,z_2^0,...,z_{n-1}^0) $. Namely, let 
$$
L(x)= \left(
\begin{array}{lr}
-i A(x)& C(x)-iB(x) \\
C(x)+iB(x) &i A(x)
\end{array} 
\right), i=\sqrt{-1}
$$
where $A(x),B(x),C(x)$ are the Jacobi polynomials (\ref{axb}). 
The spectral polynomial of $L(x)$ is $P(x,y)=y^2-f(x)$.
 Put 
$$
R(\alpha)=\left(
\begin{array}{lr}
1& \alpha \\
0 &1
\end{array} 
\right), 
R(\infty)=\left(
\begin{array}{lr}
0& 1 \\
0 &0
\end{array} 
\right) 
$$
and consider the eigenvector
 $^t(1,f_2(x,y;\alpha))$   corresponding to the
eigenvalue $y$ of the matrix $R(\alpha)L(x)R^{-1}(\alpha)$. 
This defines a divisor $D(\alpha)= (f_2)_\infty$ and hence a one-parameter
family of line bundles
$L'_{D(\alpha)}\in Pic^n (X')$, $\alpha \in \bbbp^1$ on the singular curve $X'$. 
As the line bundle
$L_{D(\alpha)}\in Pic^n (X)$ does not depend on $\alpha$ then $L'_{D(\alpha)}$,
$\alpha \in \bbbp^1$ parameterizes the fibre $\phi^{-1}(z_1^0,z_2^0,...,z_{n-1}^0) $,
 that is to say an
integral curve of (\ref{hyper}).
A simple computation shows that
$$
f_2(x,y;\alpha)=\frac
{ -iA(x)+ \alpha(C(x)+ i B(x) ) - y}
{     2i\alpha  (
-A(x)-\frac{\sqrt{-1}}{2} (\alpha-\frac{1}{\alpha})C(x)+\frac{1}{2}
(\alpha+\frac{1}{\alpha})B(x) ) }
 \; .
$$
If $D(\alpha)= \sum_{k=1}^np_k$  where $p_k=(y_k,x_k) \in X$, then
 $x_k$ is the root of the denominator
$$
-A(x)-\frac{\sqrt{-1}}{2} (\alpha-\frac{1}{\alpha})C(x)+\frac{1}{2}
(\alpha+\frac{1}{\alpha})B(x)
= -A(x)+B(x) cos(\varphi)+C(x) sin(\varphi)
$$
where  $\alpha= e^{\sqrt{-1}\varphi} $. This completes the proof of
Jacobi's theorem in the case $deg(f)=2n$.

Note that there are exactly two values $\alpha^\pm$ of $\alpha$ such that the 
pole divisor of 
$f$ is not contained in the affine part of the curve $X$ and hence
the line bundle $L'_{D(\alpha)}$ is not defined. Thus topologically
the integral curve of (\ref{hyper}) is $\bbbp-\{\alpha^+,\alpha^-\} \sim \bbbc^*$
as we explained before. At last if $deg(f)=2n-1$ Jacobi's theorem 
holds too (although Jacobi did not study this case explicitly). The differential  (\ref{diff}) is of third kind, $m=2\infty $, where
$\infty$ is its double pole, and $J(X')$ is a non-trivial extension of $J(X)$ by
$\bbbc$              
$$
0\stackrel{ }{\rightarrow} \bbbc
 \stackrel{ }{\rightarrow}
J(X')\stackrel{\phi}{\rightarrow}
J(X) \rightarrow 0 \; .
$$
Indeed, in this case $b_0^2+c_0^2-a_0^2=0$, so $\alpha^+=\alpha^-$ and
the fiber is isomorphic to $\bbbp-\{\alpha^\pm\} \sim \bbbc$.

\end{document}